# Measurements and Numerical Calculations of Thermal Conductivity to Evaluate the Quality of β-Gallium Oxide Thin Films Grown on Sapphire and Silicon Carbide by Molecular Beam Epitaxy

Diego Vaca[1], Matthew Barry[1], Luke Yates[2,b)], Neeraj Nepal[3], D. Scott Katzer[3], Brian P. Downey[3], Virginia Wheeler[3], Luke Nyakiti[4], David J. Meyer[3], Samuel Graham[5,c)], and Satish Kumar[1,a)]

[1.] School of Mechanical Engineering, Georgia Institute of Technology, Atlanta, GA 30332, USA

[2.] Sandia National Laboratories, Albuquerque, NM 87185, USA

[3.] U.S. Naval Research Laboratory, Washington, DC 20375, USA

[4.] Department of Material Science and Engineering & Marine Engineering and Technology, Texas A&M University,

College Station, TX 77843, USA

[5.] School of Engineering, University of Maryland, College Park, MD 20742, USA

*Abstract*—We report a method to obtain insights into lower thermal conductivity of β-$Ga_2O_3$ thin films grown by molecular beam epitaxy (MBE) on c-plane sapphire and 4H-SiC substrates. We compare experimental values against the numerical predictions to decipher the effect of boundary scattering and defects in thin-films. We used time domain thermoreflectance (TDTR) to perform the experiments, density functional theory and the Boltzmann transport equation for thermal conductivity calculations, and the diffuse mismatch model for TBC predictions. The experimental thermal conductivities were approximately 3 times smaller than those calculated for perfect $Ga_2O_3$ crystals of similar size. When considering the presence of grain boundaries, gallium and oxygen vacancies, and stacking faults in the calculations, the crystals that present around 1% of gallium vacancies and a density of stacking faults of $10^6$ faults/cm were the ones whose thermal conductivities were closer to the experimental results. Our analysis suggests the level of different types of defects present in the $Ga_2O_3$ crystal that could be used to improve the quality of MBE-grown samples by reducing these defects and thereby produce materials with higher thermal conductivities.

Future applications such as power electronics for AC/DC conversion or faster wireless networks will require devices with superior power density and power switching capabilities. To achieve this, semiconductors such as GaAs, SiC and GaN are commercially available and are under development to broaden their applications. On the other hand, $Ga_2O_3$, an ultra-wide bandgap semiconductor, is expected to show superior properties when used for power switching devices, according to the Baliga figure of merit, that measures the power losses in switching devices [1]. In addition, wafers of Ga2O3 are available and can be fabricated with high-volume commercial techniques such as the Czocharlaski method or the edge-defined film-fed crystal growth. However, one of the limitations for the use of $Ga_2O_3$ in high frequency and high-power switching applications is its lower thermal conductivity. The maximum reported bulk thermal conductivity of $Ga_2O_3$ at 300 °K is around 26 W/(m·K) in (010) direction [2, 3]. The low thermal conductivity can hinder the full potential of $Ga_2O_3$-based devices because high temperatures will accelerate the degradation of the device [4, 5].

Recently, many studies focused on the electro-thermal transport in $Ga_2O_3$ transistors have been reported [6-10]. In the case of thin-films transistors, the layers can be obtained using two prime methods: mechanical

---

exfoliation of the membranes or epitaxial growth on compatible substrates. In the case of exfoliation, devices with membrane thickness in the range of tens to hundreds of nm have been presented [11-13]. Even though many studies used mechanical exfoliation of $Ga_2O_3$ to fabricate devices, this process cannot be scaled for industrial production because it is neither possible to control the thickness nor the lateral dimension of the membranes. In addition, the membranes and the substrate are coupled by weak Van der Waals forces, which results in a low thermal boundary conductance (TBC). In general, the TBC at the membrane interface that have been transferred to a substrate is one order of magnitude lower than the TBC between thin layers grown or deposited using physical/chemical methods on a substrate [14, 15]. For the devices with epitaxial grown thin film layers [16-18], the thickness of layers were around 200 - 600 nm. The TBC at the interface of the $Ga_2O_3$ thin-film and substrate will be dependent on the growth process followed. Most of the previous work focuses on analyzing the electrical characteristic of the devices but only a handful of studies investigates the TBC at the interfaces of $Ga_2O_3$ thin-films and the effect of interfaces on the thermal conductivity of thin-films [14, 19].

There are multiple methods for the epitaxial growth of $Ga_2O_3$ thin layers such as low-pressure chemical vapor deposition (LPCVD), metal-organic chemical vapor deposition (MOCVD), metal-organic chemical vapor phase epitaxy (MOVPE), molecular beam epitaxy (MBE), etc. [20]. Growth of $Ga_2O_3$ thin-films on high thermal conductivity foreign substrates can provide a pathway for the development of high power devices. An interesting alternative for heteroepitaxy growth of $Ga_2O_3$ thin layers is the homoepitaxy growth on a $Ga_2O_3$-on-SiC composite wafer [21]. LPCVD and MOCVD require less expensive equipment and can have higher growth rates. The choice of growth method influences both the thermal conductivity of $Ga_2O_3$ thin layers and the TBC at their interfaces because different growth methods can create different types and concentrations of defects. For example, Song et al. [22] showed that there is a tradeoff between the quality of $Ga_2O_3$ grown on c-plane sapphire using MOVPE and the TBC. Samples grown on 6°off cut c-plane sapphire had higher thermal conductivity (10-30%), but lower TBC than those grown on 0°off cut c-plane sapphire. Sapphire has relatively lower thermal conductivity (~ 30 W/mK) and epitaxial growth on high conductivity substrates like SiC is highly desired. Even if the quality of $Ga_2O_3$ on SiC is lower than the other substrates, the thermal performance of this combination can be superior because the thermal conductivity of SiC is ~6 times higher than sapphire. MBE can be used for the heteroepitaxy growth of $Ga_2O_3$ and investigation of the fabrication of thin-films of $Ga_2O_3$ on SiC.

An important aspect to analyze for $Ga_2O_3$ grown on foreign substrates is how the polycrystalline nature of the material (grain size) along with the formation of point defects (vacancies) and linear defects (dislocations), will affect its thermal conductivity. The influence of vacancies in the crystal lattice of $Ga_2O_3$ on its thermal conductivity has been studied by incorporating the defect-induced phonon scattering rate into the solution of the Boltzmann Transport Equation (BTE) [23] and the molecular dynamic simulations [24]. However, neither the influence of linear defects nor the mean grain size of polycrystalline $Ga_2O_3$ has been considered in the previous studies along with vacancies.

We have published preliminary results on the measurement of cross-plane thermal conductivity of β-$Ga_2O_3$ and TBC at its interfaces for films grown on c-sapphire and 4H-SiC substrates using MBE [25]. In this work, we compared the measured values with theoretical results and used this comparison as a tool to estimate the defect density might have been created during growth. The measurements were performed using time domain thermoreflectance (TDTR), whereas we used the iterative solution of the Boltzmann Transport Equation (BTE) to estimate the variation of the thermal conductivity with film thickness and the effect of defects on thermal conductivity. In order to make these estimations, we have used the grain size of $Ga_2O_3$ measured using AFM and linear defect density using TEM, as input to the BTE simulations. The percentage of point defects is one unknown which is not easy to measure, and we estimated that by comparing experimental and numerical results. In summary, we propose a method to study the effects of the different types of defects on the thermal conductivity of epitaxially grown $Ga_2O_3$, which can be employed to study other materials too. Finally, the diffuse mismatch model (DMM) was used to predict TBC and better explain the experimental results.

Two thin films of undoped $Ga_2O_3$ were grown on c-sapphire and 4H-SiC by MBE at the U.S. Naval Research Laboratory (NRL). MBE is a physical-vapor epitaxial growth process in high vacuum. For $Ga_2O_3$ epitaxy, ultra-high pure elemental Ga and reactive oxygen were provided using Ga effusion cell and oxygen plasma, respectively [26]. The optimized conditions used to grow $Ga_2O_3$ on sapphire and SiC, along with further characterization of the films have been published elsewhere [25, 27]. More details about growth of $Ga_2O_3$ by MBE can be found in [28]. Also, we used bulk samples of the substrates to measure their thermal conductivities. In both cases (bulk and thin-film samples), we deposited a thin Al transducer (97 nm and 93 nm, for bulk and thin-film samples, respectively) by e-beam evaporation. The TDTR measurements were performed using a Ti:Sapphire laser (wavelength=800 nm). The frequency modulation of the pump beam can be controlled by an electro-optical modulator and is doubled using a BiBO crystal. On the surface of the sample, the pump radius was ~9.95 μm, while the probe radius was ~6.2 μm. Based on the sensitivity analysis, the data obtained at modulation frequencies of 8.8 and 11.6 Mhz were used for the samples grown on sapphire and SiC, respectively. Section A of the supplementary material has more details of the TDTR system.

High Resolution Transmission Electron Microscopy (HRTEM) performed on both samples revealed a thickness of 119.4 $\pm$ 2.8 and 81.3 $\pm$ 1.3 nm for $Ga_2O_3$ on c-sapphire and 4H-SiC substrates, respectively (see Figs. 1a and 1d). In addition, the thickness of the $Ga_2O_3$ layers was measured by spectroscopic ellipsometry and X-ray reflectometry (XRR). Both HRTEM and X-Ray Diffraction (XRD) measurements showed β-$Ga_2O_3$ of orientation (-201) for both samples. Details of the structural characterization can be found in our previous work [26]. A bright-field low- and high- resolution phase contrast TEM imaging, selected area electron diffraction (SAED) imaging and post imaging analysis using Fast Fourier transform (FFT) show that $Ga_2O_3$ thin films on sapphire and 4H-SiC substrate have a crystalline structure (Figs. 1a to 1g). Defect analysis using images of phase contrast and inverse FFT lattice images revealed that both samples have 1-dimensional type of defects (dislocations), and zero-dimensional type of defects. In brief, several TEM images are processed using the FFT to identify the regions of interest and images of those regions are converted back using the inverse FFT. Then, the linear defects are counted, and this number is divided by the area of the region of interest. More details on the method to estimate defect density can be found in [29]. In this case, the defect density was 2.5 x $10^{12}$ $cm^{-2}$ and 1.0 x $10^{12}$ $cm^{-2}$ for the $Ga_2O_3/Al_2O_3$ and $Ga_2O_3$/SiC, respectively [27]. More details on the structural characterization of the samples can be found in the Section B of the supplementary material. The TDTR setup and the data interpretation for this study has been described in [30, 31].

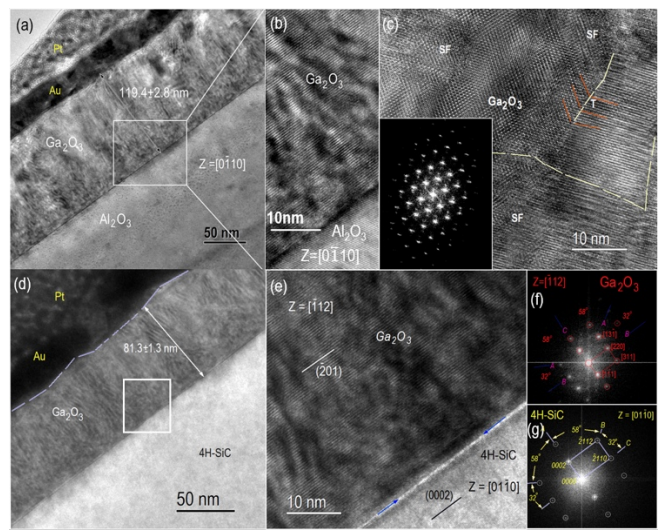

Fig. 1. A bright-field low- and high- resolution TEM micrographs showing (a-c) $Ga_2O_3$ on sapphire substrate with a thickness of ~ 119.4 nm. The presence of the lattice fringes as well as FFT reflections (inset) suggests crystalline structure. There exists lattice strain

field across the HRTEM micrograph. (c) HRTEM micrograph obtained from a specimen rotated normal to the substrates Z= [0$\bar{1}$10] direction reveal a crystalline structure decorated with high density of 1-D defects like stacking faults and twinning dislocations within the film which are indicated by SF and T, respectively. (d) bright-field TEM show ~ 81 nm thick $Ga_2O_3$/SiC. In addition, (e) HRTEM and (f and g) FFT of the film and substrate show lattice fringe real images and respective low-order diffraction reflection demonstrating crystallinity in structure of $Ga_2O_3$ film.

We used Atomic Force Microscopy (AFM) to scan the top surface of the samples and analyze the presence of lateral grains (Fig. 2). These measurements revealed that the samples are polycrystalline with a mean lateral grain size of 52 nm. This value was obtained using the intercept technique. In this technique, several random lines are drawn on the micrograph and the number of grain boundaries intersecting the lines are counted. Then the average grain size is calculating by dividing the length of the line by the number of grain boundaries. This procedure was repeated for 10 lines to have a representative value for the micrography. It is worth to mention that Fig. 2 was obtained using the "phase" signal, because this signal allows for a better contrast image that reveals the presence of grain boundaries, as opposed to the AFM image presented in [27], that used the "topographic (height)" signal, which is more suitable to measure the roughness of the sample.

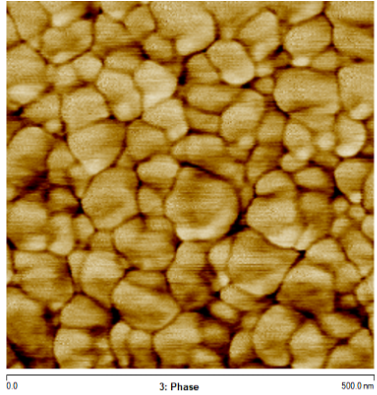

Fig. 2. AFM image of the surface of a β-$Ga_2O_3$ sample grown on SiC for this work. The mean lateral grain size is 52 nm. The field of view of the image is 500 x 500 $nm^2$

Density functional theory (DFT) calculations in conjunction with Boltzmann Transport Equations has been used for the estimation of phonon relaxation time and thermal conductivity of $Ga_2O_3$ [32]. Density functional theory (DFT) calculations were performed using the Vienna *ab initio* simulation package (VASP) to compute interatomic force constants of $Ga_2O_3$. A plane-wave basis set and the projector augmented-wave (PAW) method were used with the Perdew-Burke-Ernzerhof (PBE) exchange-correlation functional [33-36]. A 500 eV kinetic energy cutoff was used to perform the structure optimization and to calculate the second-order harmonic and third-order anharmonic force constants (IFCs). The convergence criteria for the energy and force were $10^{-9}$ eV and -0.001 eV/Å, respectively. The 20-atom β-$Ga_2O_3$ unit cell was optimized using a 4×16×8 grid for Brillouin zone sampling. The optimized lattice parameters were a = 12.45 Å, b = 3.08 Å, and c = 5.86 Å, with β = 103.76°, which are in good agreement with recent computational [23] and experimental results [37]. The second-order harmonic and third-order anharmonic IFCs were calculated using the finite displacement method with a 1×4×2 supercell of the optimized 20-atom unit cell [38]. The finite displacement distance was 0.01 Å and a $4^{th}$ nearest neighbor cutoff was used for computing the third-order IFCs. Using the second- and third-order IFCs, the phonon relaxation times and the anisotropic thermal conductivities of bulk β-$Ga_2O_3$ were calculated using Fermi's golden rule [39] with the iterative solution to the BTE [40, 41]. A 5×17×9 k-space sampling mesh was used for Brillouin zone sampling.

After calculating the IFC's, we solved the linearized form of the BTE using ShengBTE to obtain the thermal conductivity tensor (see section D of the supplementary material for details). The combined effects of sample size [42], oxygen or gallium vacancies, grain boundaries, and linear defects on the thermal conductivity were considered by adding the phonon boundary scattering rate, the vacancies-induced scattering rate, the grain boundary scattering rate, and the linear defect scattering rate (dislocations), respectively, to the anharmonic phonon scattering rate [43], according to the Matthiessen's rule:

$$1/\tau = 1/\tau_{anh} + 1/\tau_b + 1/\tau_V + 1/\tau_{gb} + 1/\tau_{ld} \quad (1)$$

Here $1/\tau_{anh}$ is the intrinsic anharmonic phonon scattering rate, $1/\tau_b$ is the phonon boundary scattering rate, $1/\tau_V$ is the phonon scattering rate due to the vacancies, $1/\tau_{gb}$ is the scattering rate due to grain boundaries, and $1/\tau_{ld}$ is linear defect scattering rate.

The phonon scattering rate caused by oxygen or gallium vacancies in the crystal can be expressed as [44, 45]:

$$1/\tau_V = x\left(-\frac{M_V}{M} - 2\right)^2 \frac{\pi}{2} \frac{\omega^2 g(\omega)}{G} \quad (2)$$

where $x$ is the density of vacancies, $M$ is the average mass per atom, $M_V$ is the mass of the missing atom, $g(\omega)$ is the phonon density of states, and $G$ is the number of atoms in the crystal (number of atoms in the unit cell). The grain boundary scattering rate was estimated using the Casimir model [46]:

$$1/\tau_{gb} = \frac{1-p(\omega)}{1+p(\omega)} D_{avg}^{-1} v_{g,a} \quad (3)$$

where $p(\omega)$ is the specularity parameter, $D_{avg}^{-1}$ is the average grain size of polycrystalline samples, and $v_{g,a}$ is the phonon group velocity along $a$ direction. In this case, the specularity parameter $p(\omega)$ was chosen to be zero, which represents diffusive scattering at grain boundaries. The linear defect scattering rate was estimated using the equation in [47]:

$$1/\tau_{ld} = 0.7 \frac{a^2}{v} \gamma^2 \omega^2 N_s \quad (4)$$

where $a$ is the lattice parameter, $v$ is the phonon velocity, $\gamma$ is the Grunnessien's constant, $\omega$ is the angular frequency and $N_s$ is the number of linear faults per cm. In contrast to the Debye-Callaway model used by other researchers [19], using an exact iterative solution of phonon BTE is more accurate, and allowed us to obtain the thermal conductivity tensor for layers with different thickness, and calculate the thermal conductivity along different directions, following the procedure presented in [48].

For the calculation of the TBC, we used the diffusive mismatch model (DMM) considering not only the acoustic phonon branches, but all branches of the materials because the complex crystalline structure leads to large number of optical branches and cannot be omitted in the calculations. For example, researchers in [23] demonstrated that, depending on the orientation of the crystal, optical phonon modes contribute significantly to the thermal conductivity of $Ga_2O_3$. We implemented the qDMM model, where the integration is performed over the wave vector $q$ [49].

Before determining the thermal conductivity of $Ga_2O_3$ and TBC at its interfaces with the substrate, we first determined the thermal properties of the sapphire and SiC substrates using 2-layered samples. The through-plane and in-plane thermal conductivities of SiC were estimated to be 301.4 ± 36.2 W/(m·K) and 387.3 ± 46.5 W/(m·K), respectively. For sapphire, the thermal conductivity was 27.3 ± 2.0 W/(m·K). These values agree with the previous studies and were used as constants in the 3-layer models. The set of data with the best fitting results were used to estimate the thermal conductivity and the TBC. The samples on c-sapphire (119 nm) had a thermal conductivity of 3.2 ± 0.3 W/(m·K), whereas the thermal conductivity of the sample on 4H-SiC (81 nm) was 3.1 ± 0.5 W/(m·K). The numerically estimated conductivities using BTE for the two thin film samples were 8.9 W/(m·K) (119 nm) and 7.9 W/(m·K) (81 nm), which are much higher than the measured values.

The thermal conductivity of crystalline samples of $Ga_2O_3$ thinner than 120 nm has been hardly reported. Thermal conductivity of thin films of $Ga_2O_3$ fabricated using PLD has been reported in [19]. The thermal conductivity of the samples fabricated by MBE are slightly higher than those fabricated by PLD with comparable thickness (~100 nm). It is lower than film grown by MOVPE on sapphire in [22], but those films were thicker (>164nm). A comparison of different results is presented in Fig. 3a. The thermal conductivity of MBE-grown samples could also be compared with single-crystal thin-film samples. For example, the study in [50] presents the thermal conductivity of monocrystalline $Ga_2O_3$ thin-films bonded to SiC. H ions were implanted in the $Ga_2O_3$ crystal before the bonding process, which might have induced strain in the crystal and produce defects. For this reason, the thermal conductivity of a ~140 nm sample was 2.9 W/(m·K), which is lower than what could be expected for a single crystal sample but comparable with our results. It is likely that the fabrication process and post-fabrication treatment will affect both the thermal conductivity and the TBC.

The phonon dispersion curve for $Ga_2O_3$ in the (-201) direction was calculated using the $2^{nd}$ order harmonic force constants (IFCs) and thermal conductivity from iterative solution of BTE using $2^{nd}$ and $3^{rd}$ order IFCs, obtained from the first principles simulations. When comparing the experimental and computational results, we observe that the experimental results in the (-201) direction are ~ 2 to 3 times smaller than the results computed from the first principles. This could indicate the presence of unavoidable imperfections during fabrication of the thin-films of $Ga_2O_3$. Fig. 3b shows the variation of the computed thermal conductivity of $Ga_2O_3$ without imperfections with respect to its thickness for the (-201) direction, where it is evident that the thermal conductivity will reach plateau, corresponding to the bulk value, similar to the results published in [19].

To explain the lower thermal conductivity measured from TDTR compared to the numerical predictions, we introduced scattering due to the vacancies of gallium and oxygen, the grain boundaries, and linear defects (stacking faults) while computing conductivities (see Eq.1). The gallium vacancies had higher impact in the reduction of the thermal conductivity because gallium is heavier than oxygen. Simulations with 3 % oxygen vacancies results in thermal conductivity of 4.2 W/(m·K) and 4.0 W/(m·K) for 119 nm and 81 nm sample respectively, which is still higher than the measured conductivity and suggests the presence of Ga vacancies in our samples. If linear defects are not considered (Fig. 4a), it was estimated that the 119 nm sample had around 3 % of Ga vacancies, whereas the 81 nm sample had around 2.5 % of Ga vacancies. Through a combination of high resolution transmission electron microscopy in combination with DFT predictions [51] and positron spectroscopy [52], previous studies determined that the most likely vacancies to occur during fabrication are Ga vacancies, for both bulk and thin films. A high density of linear defects is also probable. For this reason, three levels of linear defects were introduced in the calculations for the 120 nm sample, assuming Ga vacancies (Fig. 4b). The thermal conductivity for a sample with 1% of Ga vacancies and $10^6$ linear defects/cm was 3 W/(m·K) which is close to the measured value, as opposed to 8.9 W/(m·K) for a perfect crystal. Also, the effects of the linear defects are more pronounced when the linear density is, at least, $10^6$ defects/cm. The actual linear defects density is $1.5 \times 10^6$ defects/cm. This value was obtained by calculating the square root of $2.5 \times 10^{12}$ linear defects/cm$^2$, that is the linear defect density per unit of area previously reported [27]

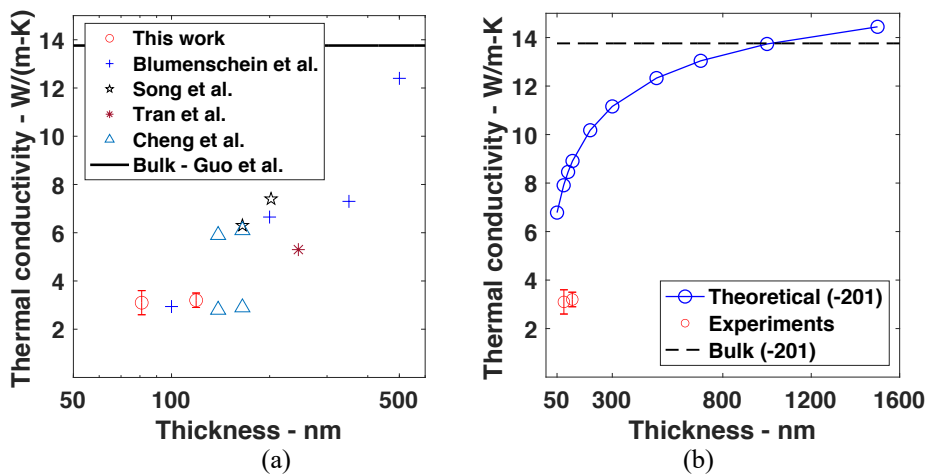

Fig. 3. (a) Comparison of published values of the thermal conductivity of thin films Ga$_2$O$_3$ with respect to their thickness. The sample for this work were grown by MBE, whereas the rest of the films were grown using PLD, MOVPE or were monocrystals [2, 19, 22, 50, 53]. (b) Variation of the thermal conductivity of Ga$_2$O$_3$ in the (-201) direction with thickness. Thermal conductivity computed using BTE is compared against the experimental results. The dashed line corresponds to the bulk thermal conductivity in the (-201) direction, as reported by [2].

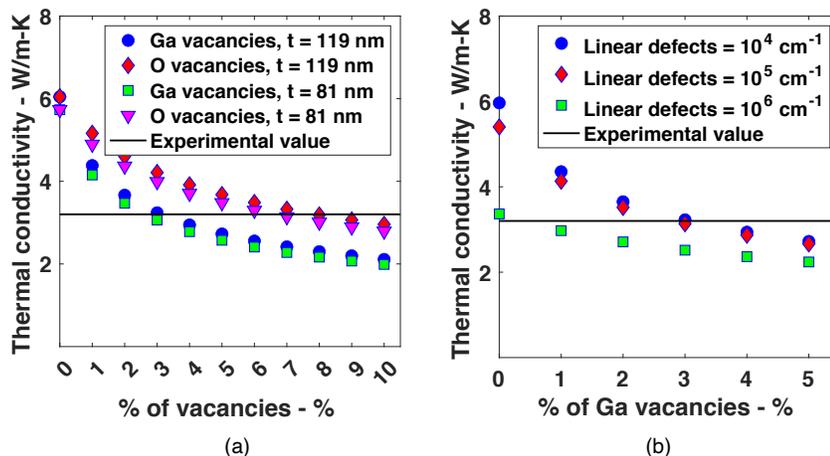

(a) (b)
Fig. 4. (a) Influence of gallium and oxygen vacancies on the thermal conductivity. Gallium vacancies cause a higher reduction in the thermal conductivity compared to oxygen vacancies. (b) Influence of the density of linear defects on the thermal conductivity when Gallium vacancies are present for the 119 nm sample (grown on sapphire).

We measured the TBC at the interface of Ga$_2$O$_3$ with the metal transducer and the substrates, SiC and sapphire. Using TDTR, the TBC at the interface of Al/Ga$_2$O$_3$ was measured as 125.41 +15.62 / -10.66 MW/(m$^2\cdot$K) for the sample grown on sapphire and 158.18 +88.37 / -41.34 MW/(m$^2\cdot$K) for the sample grown on SiC. The results may indicate that the different imperfections that the fabrication method induces in the crystal depend on the substrate itself. The TBC values reported here are in the same order of magnitude of those reported in [54-56]. For the Ga$_2$O$_3$/substrate interfaces, a summary of the TBC results is presented in Table I. Details on the methods to calculate the uncertainty can be found in the section C of the supplementary information. Due to the low parameter sensitivity, high uncertainty in the estimation of the TBC can be observed, regardless of the calculation method. Our results can be compared with TBC values presented in [50] for crystalline Ga$_2$O$_3$ films bonded to SiC. The bonding of Ga$_2$O$_3$ with SiC using surface activation bonding required a coating of thin layer of Al$_2$O$_3$ on Ga$_2$O$_3$. For a 30 nm thick Al$_2$O$_3$ layer, the reported TBC was ~ 70.1 +68.1/-10.4 MW/(m$^2\cdot$K), whereas for a 10 nm thick Al$_2$O$_3$ layer, the reported TBC was ~ 100.5 +93.8/-16.2 MW/(m$^2\cdot$K). The TBC values in our samples are higher because the addition of the Al$_2$O$_3$ layer introduces an additional resistance at the interface in the bonded samples studied in [50].

TABLE I
TBC RESULTS WITH THE UNCERTAINTY ESTIMATE WITH TWO DIFFERENT METHODS

| Interface | Traditional Method [57] | Monte Carlo Method [58] |
|---|---|---|
| Ga$_2$O$_3$/sapphire TBC MW/(m$^2\cdot$K) | 155.6 ± 65.3 | 159.6 +58.2 /-62.5 |
| Ga$_2$O$_3$/SiC TBC MW/(m$^2\cdot$K) | 141.8 ± 63.8 | 149.3 +57.9/-46.4 |

For the theoretical calculations of the TBC at the Ga$_2$O$_3$/substrate interfaces following DMM (details in the section E of the supplementary information), we used the phonon dispersion curve of Ga$_2$O$_3$ obtained from the first principles. For sapphire (orientation 001) and SiC (orientation 001), we used the dispersion curves available in the Materials Project [59, 60] in the Γ→Z and Γ→A directions, respectively. At 300 K, the calculated TBC using the DMM for the Ga$_2$O$_3$/sapphire interface was 294.3 MW/(m$^2$·K), whereas for the Ga$_2$O$_3$/SiC interface was 357.0 MW/(m$^2$·K). When comparing the experimental and theoretical results, we can see that the experimental mean TBC values are between 2 and 2.5 times smaller than the DMM based estimations. This difference could be attributed to the defects present at the interface of Ga$_2$O$_3$/substrate. One important detail in the estimation of the TBC using DMM to point out is the use of the full dispersion curves for the materials because of the complexity of the crystalline structures and the large number of optical phonons, despite their lower phonon branch velocities. If we had only considered the acoustic phonon branches, the TBC values would have been 51.8 MW/(m$^2$·K) for Ga$_2$O$_3$/sapphire and 57.4 MW/(m$^2$·K) for Ga$_2$O$_3$/SiC, underestimating the TBC values. Finally, it was suggested that Ga$_2$O$_3$ could replace GaN-based devices. Studies in [61] reported TBC values of 199.8 +29.3/-30.23 MW/(m$^2$·K) for the GaN/AlN-SiC interface and 224.41 +22.49/-23.3 MW/(m$^2$·K) for the GaN/SiC interface. Those values are between 30% and 50% higher than the ones we report for Ga$_2$O$_3$/SiC, but still lower than the TBC that could be expected for a perfect Ga$_2$O$_3$/SiC interface.

We report the thermal conductivity of Ga$_2$O$_3$ thin films grown by MBE with thicknesses of 119 nm and 81 nm, and the TBC at Ga$_2$O$_3$/sapphire and Ga$_2$O$_3$/SiC interfaces. The measured thermal conductivity is around three times smaller than the numerically calculated conductivity for pristine thin-films of similar thickness with no defects. Calculations of the variation in thermal conductivity with the percentage of vacancies of gallium and oxygen atoms, linear defects in thin-films, and the lateral grain boundaries explains the experimental results. For example, inclusion of grain boundary scattering, corresponding to grain sizes obtained from AFM, contributed to a reduction of 32% in thermal conductivity compared to a sample with no defects. 1% of Ga vacancies contributed to a further reduction of 28%, and the presence 10$^6$ cm$^{-1}$ linear defects reduced another 30% the thermal conductivity, in the samples grown on sapphire. In all, our results provide reference values of thermal properties of thin-film Ga$_2$O$_3$ and its interfaces and suggests the level of defects present in the crystal that could be used to accelerate the design of Ga$_2$O$_3$-based electronic devices.

## AUTHOR DECLARATIONS

### Conflict of interest
The authors declare no conflicts of interest

### Credits
The following article has been submitted to Applied Physics Letters. After it is published, it will be found at https://publishing.aip.org/resources/librarians/products/journals/.

## DATA AVAILABILITY

All data needed to evaluate the conclusions of this paper are present in the paper and/or the Supplementary Materials. Additional data related to this paper may be requested from the authors.

# Measurements and Numerical Calculations of Thermal Conductivity to Evaluate the Quality of β-Gallium Oxide Thin Films Grown on Sapphire and Silicon Carbide by Molecular Beam Epitaxy

Diego Vaca[1], Matthew Barry[1], Luke Yates[2,b)], Neeraj Nepal[3], D. Scott Katzer[3], Brian P. Downey[3], Virginia Wheeler[3], Luke Nyakiti[4], David J. Meyer[3], Samuel Graham[5,c)], and Satish Kumar[1,a)]

[1.] School of Mechanical Engineering, Georgia Institute of Technology, Atlanta, GA 30332, USA
[2.] Sandia National Laboratories, Albuquerque, NM 87185, USA
[3.] U.S. Naval Research Laboratory, Washington, DC 20375, USA
[4.] Department of Material Science and Engineering & Marine Engineering and Technology, Texas A&M University, College Station, TX 77843, USA
[5.] School of Engineering, University of Maryland, College Park, MD 20742, USA

## Supplementary A: Details on the TDTR system

This information has been taken from [1]. The TDTR two-color system consist of a Spectra Physics Ti:Sapphire (λ=800 nm, 40 nJ/pulse) laser with ~150 fs pulse width and a repetition rate of ~80 MHz. The beam is split in a probe and a pump beam. The frequency modulation of the pump beam can be controlled by an electro-optical modulator, and is doubled using a BiBO crystal. The probe beam enters a double-pass delay stage before merging its path again with the pump beam. Both collinear beams were focused by a 10X objective lens on the sample. On the surface of the sample, the pump radius was ~9.95 μm, while the probe radius was ~6.2 μm. The TDTR setup used in this work has available six filters that correspond to 1.2, 2.2, 3.6, 6.3, 8.8 and 11.6 MHz for the modulation frequency. Based on the sensitivity analysis, the data obtained at modulation frequencies of 8.8 and 11.6 Mhz were used for the samples grown on sapphire and SiC, respectively.

For the thermal properties of the susbtrates, we used 2-layered samples to measure these properties. In the case of SiC the through-plane and in-plane thermal conductivities were estimated to be 301.4 ± 36.2 W/m-K and 387.3 ± 46.5 W/m-K, respectively. In the case of sapphire the thermal conductivity was measured as 27.3 ± 2.0 W/m-K. These thermal conductivity values are in the range of the values published in [2, 3]. In addition, the values for the heat capacity were taken from published literature and were 778.8 J/kg-K, 669.3 J/kg-K and 484.8 J/kg-K for sapphire [4], SiC [5] and $Ga_2O_3$ [6], respectively. Finally, the thermal conductivity of a thin film Al transducer was estimated using the four-point collinear probe method to measure the electrical sheet resistivity and converting that to thermal conductivity using the Wiedemann-Franz Law, and it was measured as 170 W/m-K.

All measurements were performed in the Prof. Samuel Graham's laboratory by Diego Vaca and Luke Yates.

## Supplementary B: Details on the structural characterization of the samples

Raman spectroscopy was used to study the crystallinity of $Ga_2O_3$ thin-films. The measurements were performed on the samples (Porto's notation) [7] using a Thermo Scientific DXR Raman system. 532 nm laser of 9 mW optical power was focused to a spot size of 1.1 μm on the sample using a 50x objective and the signal was collected using the same objective. The Raman spectra (Fig. A1) shows that the film has monoclinic structure. $A^3_g$ mode represents the translation of tetrahedral and octahedral chains, whereas $A^{10}_g$ represents bending and stretching of $GaO_4$ tetrahedral [8, 9].

In addition, to evaluate the quality of the polycrystalline samples, we compared the FWHM of the samples with other homo-epitaxially and hetero-epitaxyally grown thin films. Researchers in [8] reported 40 arc-sec FWHM for a thin film grown on $Ga_2O_3$ by low-pressure CVD, whereas [10] reported 1.54 degrees FWHM and 3.09 degrees FWHM for samples grown on c-plane sapphire by MOCVD and low-pressure CVD, respectively. Our samples are 694 arc-sec (0.19 degrees) FWHM, which indicates low mosaicity (i.e., good crystallinity) [11]

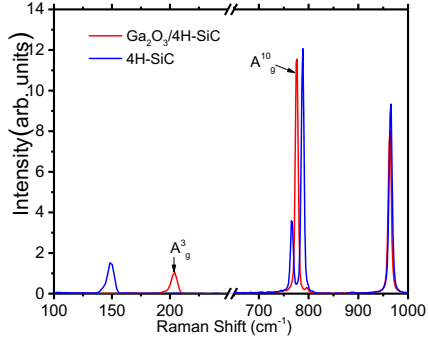

Fig. A1. Raman Spectra of β-Ga$_2$O$_3$/4H-SiC and 4H-SiC.

**Supplementary C: Methods to calculate uncertainty**

One of the challenges while using TDTR is the selection of the appropriate experimental conditions, considering sensitivity with respect to different parameters and how they affect the determination of unknown parameters, i.e., the thermal conductivity and the TBC for the present analysis. More details about the sensitivity analysis of TDTR can be found in our previous publication [23]. For the case of thermal conductivity, the sensitivity is high enough to conclude that the experimental conditions used in the measurements were adequate to estimate an accurate value. However, the sensitivity to TBC is low for these experimental conditions. Because of the low sensitivity of the TBC values, the uncertainty of those measurements was estimated using two methods. The first method is a traditional method which uses a mathematical expression employed in [12]. The second method was a Monte Carlo method [13], which uses a population of 1000 data sets, randomly chosen from a normal distribution. Each data set is fitted with a different group of model parameters to generate a random distribution of possible results. Once the distribution of possible results is obtained, the value that is at the 50$^{th}$ percentile of the distribution is taken as the mean value, whereas the values at the 5$^{th}$ percentile and the 95$^{th}$ percentile were the lower and upper bounds, respectively.

**Supplementary D: Calculations of the lattice thermal conductivity in the (-201) direction**

The procedure to calculate the thermal conductivity is described in [14]. In summary, after calculating the IFC's (see section C), we solved the linearized form of the Boltzmann Transport Equation (BTE) using ShengBTE [15], that requires sets of second-order and third-order force constants. ShengBTE uses Eqs. D1a and D1b to calculate the thermal conductivity tensor

$$\kappa_l^{\alpha\beta} = \frac{1}{k_B T^2 \Omega N} \sum_\lambda f_o (f_o + 1)(\hbar\omega_\lambda)^2 v_\lambda^\alpha F_\lambda^\beta \qquad \text{(D1a)}$$

$$F_\lambda = \tau_\lambda^0 (v_\lambda + \Delta_\lambda) \qquad \text{(D1b)}$$

where $\kappa$ is the component of the tensor of the thermal conductivity in the cartesian direction $\alpha$ and $\beta$, $k_B$ is the Boltzmann constant, $T$ is the temperature, $\Omega$ is the volume of the unit cell, $N$ is the number of discrete points of a regular grid in the Brillouin zone ($N = N_1 \times N_2 \times N_3$ q points with q being the wave vector), $\lambda$ is the phonon mode, $f_o$ is the equilibrium Bose-Einstein, $\hbar$ is de modified Plank constant, $\omega$ is the frequency, $v$ is the group velocity, $F$ is a linearized form of the BTE, $\tau$ is the relaxation time, $v$ is the group velocity and $\Delta$ is a term that measures how much or the population of a specific phonon mode deviates from the relaxation time approximation.

The software calculates the thermal conductivity tensor for the Ga$_2$O$_3$ crystal structure in this form:

$$\kappa_{ij} = \begin{pmatrix} \kappa_{xx} & \kappa_{xy} & \kappa_{xz} \\ \kappa_{yx} & \kappa_{yy} & \kappa_{yz} \\ \kappa_{zx} & \kappa_{zy} & \kappa_{zz} \end{pmatrix} \tag{D2}$$

The directions (100) and (010) coincide with $\hat{x}$ and $\hat{y}$, respectively, whereas (001) is at an angle β with respect to $\hat{x}$ and orthogonal to $\hat{y}$.

For example, the tensor that we obtained for the 81 nm-thick samples was (units are W/m-K):

$$\kappa_{ij} = \begin{pmatrix} 5.94 & 0 & 0.14 \\ 0 & 12.93 & 0 \\ 14.33 & 0 & 8.06 \end{pmatrix}$$

For the calculation of the thermal conductivity in the (-201) direction we used Eq. D3

$$\kappa_{(-201)} = K_{xx} \sin^2\gamma + K_{xz} \sin 2\gamma + K_{zz} \cos^2\gamma \tag{D3}$$

where γ = 167.7 °, that corresponds to the angle between $\hat{x}$ and (-201)

Using Eq. D3 and the tensor for the 81 nm-thick sample, the thermal conductivity we calculated in the (-201) direction was 7.9 W/m-K. The same approach was followed for the 119 nm-thick sample.

For the calculation of the reduced thermal conductivity as a consequence of the defects, we introduced the respective scattering rates before obtaining the thermal conductivity tensor.

**Supplementary E: Calculations of the DMM model**

For the calculation of the thermal boundary conductance for energy moving from material 1 (Ga$_2$O$_3$) to material 2 (substrate), $h_{BD}^{1\to 2}$, we employed the approach found in [16] (Eq. E1), where the integration is done in the wave vector space.

$$h_{BD}^{1\to 2} = \frac{1}{8\pi^2} \sum_j \int_{q_{j,1}} \hbar\omega_{j,1}(q) q_{j,1}^2 |v_{j,1}(q_{j,1})| \frac{df_o}{dT} \zeta^{1\to 2} dq_{j,1} \tag{E1}$$

Where the summation is performed over the all phonon branches, $j$, $\hbar$ is de modified Plank constant, $f_o$ is the equilibrium Bose-Einstein distribution, $q$ is the wave vector, $\omega$ is the frequency, $v$ is the group velocity defined as $d\omega/dq$, and $\zeta^{1\to 2}$ is the transmission coefficient from material 1 to material 2. The transmission coefficient was calculated with Eq. E2.

$$\zeta^{1\to 2}(T) = \frac{\sum_j \int_{q_{j,2}} \hbar\omega_{j,2}(q) q_{j,2}^2 |v_{j,2}(q_{j,2})| f_o dq_{j,2}}{\sum_j \int_{q_{j,2}} \hbar\omega_{j,2}(q) q_{j,2}^2 |v_{j,2}(q_{j,2})| f_o dq_{j,2} + \sum_j \int_{q_{j,1}} \hbar\omega_{j,1}(q) q_{j,1}^2 |v_{j,1}(q_{j,1})| f_o dq_{j,1}} \tag{E2}$$

Equations E1 and E2 require the knowledge of the phonon dispersion curves of each material. These curves where obtained from first principles calculations in the case of Ga$_2$O$_3$ and from the Materials Project web site, in the case of SiC and sapphire. The discrete phonon dispersion curves where fit to a 4$^{th}$ order polynomial equation for each phonon branch, and these relations were used to obtain all the inputs needed to calculate the TBC.